\documentclass[aps,prb,showpacs]{revtex4}
\usepackage{latexsym}
\usepackage{amssymb}
\usepackage{graphicx}
\begin{document}
\title{Low-temperature transport through a quantum dot between two superconductor leads}
\author{S. Y. Liu}
\email{liusy@mail.sjtu.edu.cn}
\author{X. L. Lei}
\affiliation{Department of Physics, Shanghai Jiaotong University, 1954 Huashan Road, Shanghai 200030, China}
\date{\today}

\begin{abstract}
We consider a quantum dot coupled to two BCS superconductors with same gap energies $\Delta$. The
transport properties are investigated by means of infinite-$U$ noncrossing approximation. In
equilibrium density of states, Kondo effect shows up as two sharp peaks around the gap bounds.
Application of a finite voltage bias leads these peaks to split, leaving suppressed peaks near the
edges of energy gap of each lead. The clearest signatures of the Kondo effect in transport are
three peaks in the nonlinear differential conductance: one around zero bias, another two at biases
$\pm 2\Delta$. This result is consistent with recent experiment. We also predict that with
decreasing temperature, the differential conductances at biases $\pm 2\Delta$ anomalously increase,
while the linear conductance descends.
\end{abstract}

\maketitle

\section{Introduction}

In recent years, the Kondo effect for various mesoscopic systems has been extensively studied in
literature. Most investigations have been concerned with a quantum dot between two normal metallic
leads.\cite{Kondo} In density of states, it has been found an emergence of a sharp Kondo peak near
Fermi surface\cite{Spectrum} and a split of this peak when out of equilibrium.\cite{Wingreen} In
the voltage-dependent differential conductance, there appears an anomalous zero-bias maximum,
ascending with a decrease of temperature.\cite{Kondo,Wingreen,Hettler} The further studies
indicated that the manifestation of the Kondo effect is strongly dependent on the properties of
leads, by which the quantum dot is sandwiched. Since in BCS superconductors there are present
energy gaps, it is very interesting what the Kondo effect exhibits in the case of two
superconductor leads. How are transport properties modified?

For a quantum dot between two superconductors, recent studies have shown that the Kondo effect will
cause two sharp peaks near gap edges in the equilibrium density of
states.\cite{Ishizaka,Krawiec1,Choi} This conclusion agrees with the investigation for dilute
magnetic alloy superconductors.\cite{Chung} In superconductors, all conduction electrons within
energy gap are paired into Cooper singlets and the Kondo-correlated states only can form around the
energy-gap bounds. In addition, since the leads coupled to a quantum dot are easily biased to
nonequilibrium, it is of course raised the question, out of equilibrium, how the Kondo effect
affects on the density of states. Unfortunately, until now, the study about this subject has not
been carried out.

In theory, the transport properties have not been well studied yet. Ivanov predicted two surprising
maxima in current near biases $\pm 2\Delta$.\cite{Ivanov} Moreover, Golub {\it et al.} only carried
out the current for voltage bias in subgap regime.\cite{Golub} Very recently, the nonlinear
transport through a quantum dot connected with two superconductors has been reported in experiment
and rich phenomena were observed.\cite{Buitelaar} In the Kondo regime, except the peaks coming from
Andreev reflection, it has been found two peaks at biases $\pm 2\Delta$ and one peak at zero-bias
in the differential conductance. The last one has been explained as the result of Kondo effect and
the nature of the remaining two peaks stayed unclear, however.

In present paper, we use the infinite-$U$ noncrossing approximation (NCA) to investigate the
nonequilibrium Kondo effect for a quantum dot coupled to two superconductors. It is well known, to
study the Kondo problem, NCA is a powerful tool and even can yield quantitative
results.\cite{Wingreen,Hettler,Bickers} We find that a voltage bias between two superconducting
leads will cause the equilibrium Kondo peaks in density of states to split. The resulting peaks
appear at the edges of energy gap of each lead and their amplitudes are suppressed. In the
nonequilibrium differential conductance versus bias, there are observed two anomalous peaks at
biases $\pm 2\Delta$ and a small peak at zero bias. Our study indicates that the peaks at biases
$\pm 2\Delta$, observed in experiment,\cite{Buitelaar} also come from the Kondo effect.

This paper is organized as follows. In section II we introduce the model and NCA for the case of
the leads having energy gaps. In section III the numerical results are discussed. Finally, the
conclusions are present in section IV.

\section{Model and Formulation}

We consider a quantum dot coupled to two (left and right) BCS-superconducting leads. Between lead
and quantum dot a dc voltage $eV_\alpha$ ($\alpha$=L or R) is applied. In order to clearly explore
the effect of energy gaps on the Kondo resonance, here, we restrict ourself to the case where the
superconducting gap energy $\Delta$ is much larger than Kondo temperature $T_{\rm K}$. Hence, the
influence of the coupling between leads and the quantum dot on the superconductivity can be
neglected.

After performed a gauge transformation\cite{Rogovin}, the Hamiltonian of the system can be written
as
\begin{equation}
H=\sum_{\alpha=L,R} H_{\alpha} +H_{dot}+\sum_{\alpha=L,R} H_T^\alpha(\tau).
\end{equation}
Here $H_{\alpha}$ is the Hamiltonian of the superconducting lead $\alpha$,
\begin{equation}
H_\alpha=\sum_{{\bf k} \sigma}\varepsilon_{\alpha {\bf k} \sigma}^0c_{\alpha {\bf k} \sigma}^+
c_{\alpha {\bf k} \sigma}+\sum_{{\bf k}}(\Delta_{\alpha}^*c_{\alpha {\bf k} \downarrow}
c_{\alpha -{\bf k} \uparrow}+\Delta_{\alpha}c_{\alpha -{\bf k} \uparrow}^+
c_{\alpha {\bf k} \downarrow}^+),
\end{equation}
with $c_{\alpha {\bf k} \sigma}^+$ ($c_{\alpha {\bf k} \sigma}$) creating (destroying) an electron
with momentum ${\bf k}$ and spin $\sigma$ in the lead $\alpha$. The dot is modelled as an
infinite-$U$ Anderson impurity having bare energy $\varepsilon_d$. Its Hamiltonian in the
slave-boson representation is given by
\begin{equation}
H_{dot}=\varepsilon_d \sum_\sigma f_\sigma^+ f_\sigma.
\end{equation}
The $U\rightarrow\infty$ constraint of single occupancy takes the form $Q=\sum_\sigma f^+_\sigma
f_\sigma+b^+b=1$. An auxiliary boson operator $b^+$ and fermion operator $f^+_\sigma$ create an
empty site and singly occupied state, correspondingly. They relate to the ordinary electron
creation (annihilation) operator $d_\sigma^+$ ($d_\sigma$) in quantum dot by
\begin{eqnarray}
d_\sigma&=&b^+f_\sigma,\\
d_\sigma^+&=&f_\sigma^+ b.
\end{eqnarray}
The hopping between the lead $\alpha$ and the dot is described by $H_T^\alpha$,
\begin{equation}
H_T^\alpha(\tau)=\sum_{{\bf k}\sigma}V_{\alpha {\bf k}\sigma}{\rm e}^{{\rm i} \phi_\alpha}
c_{\alpha {\bf k}\sigma}^+ b^+f_\sigma +{\rm H.c.}
\end{equation}
with $\phi_\alpha= e V_\alpha\tau/\hbar$.

The NCA is based on a self-consistent Feynman propagator expansion.\cite{Bickers,Wingreen,Hettler}
To make the standard diagram techniques valid, for example the Wick's theorem, the slave boson
perturbation expansion is initially formulated in the grand canonical ensemble, i.e. in the
enlarged Hilbert space of pseudo-fermion and slave boson degrees of freedom. In a second step, the
exact projection of the equations onto the physical Hilbert space, $Q=1$, is performed.

The noncrossing approximation is accurate enough to study the dynamics of the quantum dot between
two normal leads. In this approach, only the lowest-order diagrams for the boson and fermion
propagators need to be considered. When one or both leads are replaced by superconductors,
additional diagrams should be taken into account to include the effect of multi-Andreev
reflection.\cite{Clerk} In our study, we restrict ourself to the case where the Coulomb interaction
$U$ in quantum dot is much larger than the gap energy $\Delta$ and the coupling constant between
the dot and leads $\Gamma$, i.e. in the infinite $U$ limit. Due to the energetically unfavorable of
the $2e$-charge fluctuation, the contribution of the Andreev reflection can be
neglected.\cite{Kang} Hence, the usual noncrossing approximation also can be used here.

The resultant equations for the self-energies of the retarded Green functions of the
pseudo-fermions, $G_{f\sigma}^r(\omega)=(\omega-\varepsilon_d-\Sigma_{f\sigma}^r(\omega))^{-1}$ and
the slave-bosons, $D^r(\omega)=(\omega-\Pi^r(\omega))^{-1}$, constrained to the physical subspace,
can be written as
\begin{equation}
\Sigma_{f \sigma}^r(\omega)=\sum_\alpha\int \frac{{\rm d}\varepsilon}{\pi}
\Gamma_{\alpha\sigma}(\omega-\varepsilon+eV_\alpha)(1-f(\omega-\varepsilon+eV_\alpha))D^r(\varepsilon),
\end{equation}
\begin{equation}
\Pi^r(\omega)=\sum_{\sigma\alpha}\int \frac{{\rm d}\varepsilon}{\pi}
\Gamma_{\alpha\sigma}(\varepsilon-\omega-eV_\alpha)f(\varepsilon-\omega-eV_\alpha)G^r_{f\sigma}(\varepsilon).
\end{equation}
Here, $\Gamma_{\alpha\sigma}(\omega)=\Gamma_{\alpha\sigma}\rho^{(s)}_\alpha (\omega)$. The
normalized density of states of electrons in superconductors $\rho^{(s)}_\alpha (\omega)$ has the
form \cite{Lin}
\begin{equation}
\rho^{(s)}_\alpha (\omega)={\rm Re}\frac{|\omega|}{\sqrt{\omega^2-\Delta_\alpha^2}}.\label{eq1}
\end{equation}
Coupling constant $\Gamma_{\alpha\sigma}$ is defined as $\Gamma_{\alpha\sigma}=\pi|V_{\alpha k
\sigma}|N_\alpha (0)$ with $N_\alpha (0)$ being the normal density of states in the Fermi level.

Noted that the density of states of electrons in superconductors can be written as Eq. (\ref{eq1}),
only when the gap energy $\Delta$ is much larger than the Kondo temperature. In another case, it
needs to be determined self-consistently, as shown in Ref.\,\onlinecite{Chung} for a magnetic
impurity in superconductor.

In the nonequilibrium situation, additional Green's functions named less Green's functions of
pseudo-fermions $G_{f\sigma}^<(\omega)$, and slave bosons $D^<(\omega)$, should be considered.
Standard manipulation of the nonequilibrium Dyson equations leads to
\begin{equation}
G_{f\sigma}^<(\omega)=G_{f\sigma}^r(\omega)\Sigma_{f\sigma}^<(\omega)G_{f\sigma}^a(\omega),
\end{equation}
\begin{equation}
D^<(\omega)=D^r(\omega)\Pi^<(\omega)D^a(\omega),
\end{equation}
where the self-energies are given by
\begin{equation}
\Sigma_{f \sigma}^<(\omega)=-\sum_\alpha\int \frac{{\rm d}\varepsilon}{\pi}
\Gamma_{\alpha\sigma}(\varepsilon-\omega+eV_\alpha)f(\omega-\varepsilon+eV_\alpha)D^<(\varepsilon),
\end{equation}
\begin{equation}
\Pi^<(\omega)=-\sum_{\sigma\alpha}\int \frac{{\rm d}\varepsilon}{\pi}
\Gamma_{\alpha\sigma}(\varepsilon-\omega-eV_\alpha)(1-f(\varepsilon-\omega-eV_\alpha))G^<_{f\sigma}(\varepsilon).
\end{equation}

The physical impurity Green's function is then obtained from
\begin{equation}
G_{d\sigma}^r (\omega)=\frac{1}{Z}\int \frac{d\varepsilon}{\pi}
[{\rm Im}D^r(\varepsilon)
G_{f\sigma}^<(\omega+\varepsilon)-D^<(\varepsilon){\rm
Im}G_{f\sigma}^r (\omega+\varepsilon)]
\end{equation}
with the partition function
\begin{equation}
Z=\frac{{\rm i}}{2\pi}\int d\omega [D^<(\omega)-\sum_\sigma G^<_{f\sigma}(\omega)].
\end{equation}

\section{Numerical Results}

We have numerically solved the self-consistent NCA equations. For the evaluations, it is assumed a
Gaussian density of states of normal conduction electrons with half width $D$. The single level of
the quantum dot is chosen to be $\varepsilon_d=-0.5D$. We only treat with a symmetric structure:
two s-wave superconducting leads, characterized by $\Delta=\Delta_L=\Delta_R=0.25D$, are coupled to
a quantum dot with coupling constant $\Gamma=\Gamma_{L\sigma}=\Gamma_{R\sigma}=0.125D$. In this
case the Kondo temperature for normal leads $T_{\rm K}$ is estimated to be $0.017D$.

It is well known that, for a quantum dot between two metallic leads, the set of NCA equations can
yield correct sharp peak at the chemical potential in the equilibrium density of states. This peak
at zero temperature should be understood as the exhibition of the transition from the $N$-particle
ground state to the ground state with $N+1$ or $N-1$ electrons.\cite{Wingreen} These $N\rightarrow
N+1$ and $N\rightarrow N-1$ transitions occur only when the correlated ground state has a finite
amplitude to have an empty site. Since by definition the ground state energies differ by the
chemical potential, the Kondo peak in density of states should be formed at the Fermi level.

\begin{figure}
\includegraphics [width=0.45\textwidth,clip] {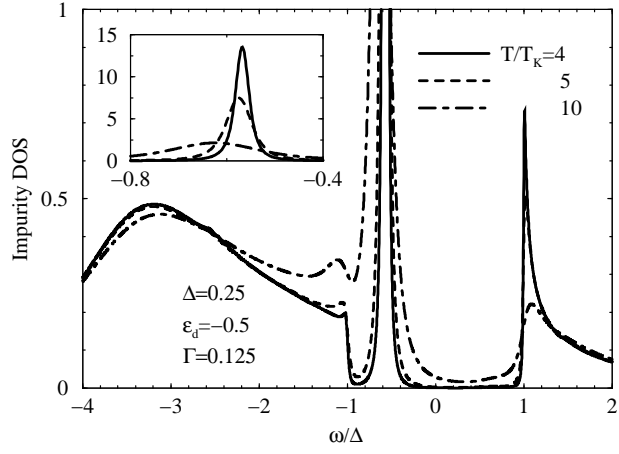}
\caption{The equilibrium density of states for a quantum dot coupled to two superconducting leads
at several temperatures: $T=10$\,$T_{\rm K}$, $5$\,$T_{\rm K}$ and $4$\,$T_{\rm K}$. The inset
shows a blowup of the region around $\omega=-\Delta$.} \label{fig1}
\end{figure}
When the leads become superconductors with $\Delta\gg T_{\rm K}$, all electrons within the gap are
paired into Cooper singlets. The occurrence of the $N\rightarrow N+1$ and $N\rightarrow N-1$
transitions should cost additional energy of order $\pm\Delta$. Consequently, in the impurity
density of states, the sharp Kondo peaks are expected to be observed near the gap bounds. Indeed,
from Fig.\,1, these two peaks at $\pm \Delta$ are clearly visible and a gap opens around Fermi
surface. We also can see that the sharp peak at $-\Delta$ shifts towards the center of gap. The
shape of this peak is more sharp and its amplitude is more large, by comparison with the peak at
$\Delta$. When temperature decreases, both peaks become more pronounced. The small weights of these
two peaks reflect the fact that the probability of site being unoccupied in the ground state is
small. Noted that with lower temperature, the shape of the peak around $-\Delta$ is very sharp and
the numerical calculation is difficult to be carried out.

In the impurity density of states, there is another low, broad peak around bare-level energy. This
peak is associated with the transition of the empty site to excited states. For finite interaction
energy $U$, an additional broad peak in the density of states near $\varepsilon_d+U$ should appear.
At the same time, the peak around $\Delta$ also will move towards the center of gap, as that shown
in Ref.\,\onlinecite{Ishizaka}.

\begin{figure}
\includegraphics [width=0.45\textwidth,clip] {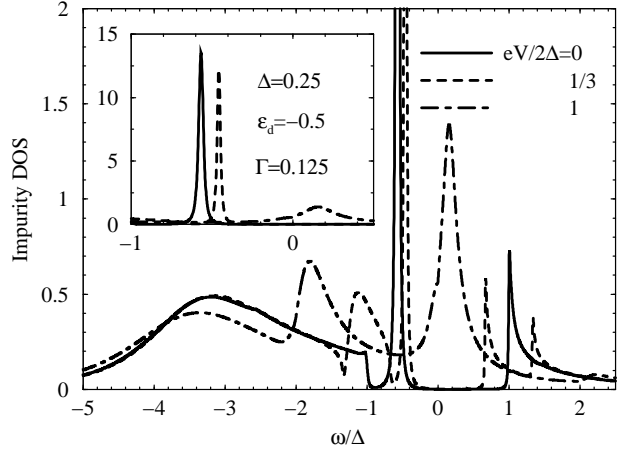}
\caption{The nonequilibrium density of states for a quantum dot between two superconducting leads.
Several biases are applied: $eV/2=0$, $\Delta/3$ and $\Delta$. The temperature is $T=4$\,$T_{\rm
K}$. The inset shows a blowup of the region around $\omega=-\Delta$. } \label{fig2}
\end{figure}
We also have carried out the nonequilibrium density of states of impurity. The results are plotted
in Fig.\,2. It can be seen that when a voltage bias $eV_L=-eV_R=eV/2$ is applied, the equilibrium
peaks at $\pm\Delta$ split and the resulting peaks are formed around $\pm eV/2\pm \Delta$. In
comparison with the corresponding equilibrium peaks, the amplitudes of these split peaks decrease
and their shapes become more broadening. For the bias $eV/2=\Delta$, there only three peaks are
observed, however. The split peaks near $eV/2-\Delta$ and $-eV/2+\Delta$ are superposed and a more
broadening peak appears. Noted that, the positions $eV/2-\Delta$ and $-eV/2+\Delta$ also correspond
to the lower edge of left-lead gap and upper edge of right-lead gap, respectively.

Physically, out of equilibrium, the quantum fluctuation will produce a finite probability of an
empty site. Hence, the Kondo peaks in density of states also can occur at the two edges of energy
gap of each lead, i.e. near $\pm \Delta \pm eV/2$.

We also have carried out nonlinear differential conductance after computing the current by formula
\begin{equation}
I_L(V)=-\frac{e}{\hbar}\sum_{\sigma}\int d\omega
\Gamma_{L\sigma}(\omega-\mu_L)[G_{d\sigma}^<(\omega)-{\rm
Im}G_{d\sigma}^{(r)}(\omega)f(\omega-\mu_L)],
\end{equation}

\begin{equation}
I_R(V)=\frac{e}{\hbar}\sum_{\sigma}\int d\omega
\Gamma_{R\sigma}(\omega-\mu_R)[G_{d\sigma}^<(\omega)-{\rm
Im}G_{d\sigma}^{(r)}(\omega)f(\omega-\mu_R)],
\end{equation}
where the lesser Green function of the impurity $G_d^<$ can be obtained from the pseudo-fermion and
slave boson Green functions via
\begin{equation}
G_{d\sigma}^<(\omega)=\frac{1}{Z}\int
\frac{d\varepsilon}{2\pi}G_{f\sigma}^<(\varepsilon)D^r(\varepsilon-\omega).
\end{equation}
Our study is limited to the case of infinite $U$. The contribution of Josephson effect to current
is small and can be neglected.\cite{Ishizaka}

Since the NCA is a conserving approximation, the currents computed from the left and the right
leads should be the same. In calculation, the current conservation within the NCA has been checked
and the agreement between two currents within $0.5\%$ is found. We evaluate the differential
conductance, $G(V)=dI(V)/dV$, by numerical derivative.

\begin{figure}
\includegraphics [width=0.45\textwidth,clip] {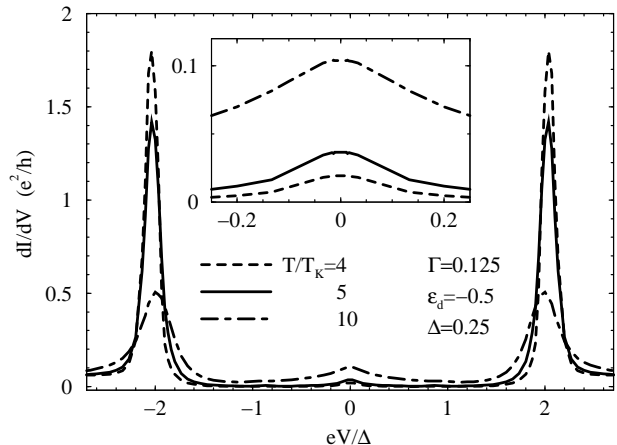}
\caption{The voltage dependence of differential conductance for a quantum dot connected with two
superconductors. The inset shows a blowup of the region around zero bias.} \label{fig3}
\end{figure}
We plot the nonlinear differential conductance versus bias in Fig.\,3 for several temperatures. A
small peak at zero bias always can be seen. In contrast to the case of normal leads, this zero-bias
maximum descends with decreasing the temperature. It is also observed two anomalous peaks in
differential conductance at biases $eV=\pm 2\Delta$. When temperature falls, the shapes of these
two peaks become more sharp and their amplitudes increase. In the following, we will clarify that
all these three peaks are produced by Kondo effect.

To explain the absence of anomaly of zero-bias conductance, we rewrite the linear differential
conductance as
\begin{equation}
\frac{dI}{dV}(V=0)=\frac{e^2}{h}\sum_{\alpha\sigma}\Gamma_{\alpha\sigma}\int d\omega
\rho^{(s)}_\alpha (\omega)\left ( -\frac{\partial f(\omega)}{\partial \omega}\right ){\rm
Im}G_{d\sigma}(\omega).
\end{equation}
Since $\rho^{(s)}_\alpha (\omega)$ only has a nonzero value when out of energy gap, the
contributions of the impurity spectrum and the Fermi-function differential to conductance come from
the energy regime where $|\omega|>\Delta$. The lower-energy Kondo peak in equilibrium density of
states appears within energy gap (see Fig. \,1), so its contribution can be neglected. When the
temperature falls, the higher-energy Kondo peak increases slowly and the exponential diminution of
Fermi-function differential becomes dominant. In results, linear differential conductance is
suppressed. Noted that the absence of anomaly of zero-bias conductance also has been carried out
for a quantum dot connected with a metallic and a superconductor lead.\cite{Krawiec}

When the bias deviates from zero, splitting of equilibrium Kondo peaks in density of states and
suppression of the resultant peaks in amplitudes should cause a decrease of the differential
conductance. However, for $|eV|\approx 2\Delta$, from Fig.\,2 we can see, in the density of states,
there appears a peak produced by superposition of the higher-energy and lower-energy Kondo peaks.
The position of this peak coincides with the edges of left and right leads, so perfect transparency
occurs. Hence, in differential conductance, two peaks around biases $\pm 2\Delta$ emerge.
Furthermore, with decreasing temperature, the enhancement of the amplitude of this superposition
peak leads to a rapid increase of nonlinear conductance near biases $\pm 2 \Delta$.

In the experiment, besides the peaks induced by Andreev reflection, such three peaks in voltage
dependence of nonlinear conductance also have been observed.\cite{Buitelaar} However, there, the
origin of the peaks at $\pm 2\Delta$ was not clear. Our calculation indicated these two peaks are
produced by Kondo effect. Unfortunately, the quantitative comparison between our results and
experiment can not be made, since our study for $\Delta \gg T_{\rm K}$ corresponds to the
experimental situations, where have two energy levels of quantum dot down to Fermi surface.

\section{Conclusions}
We have studied the nonequilibrium Kondo effect for a quantum dot coupled to two superconductors
with energy gap $\Delta\gg T_{\rm K}$. Employing the infinite-$U$ noncrossing approximation, we
have carried out the nonequilibrium impurity density of states.  When a voltage bias is applied to
the both side of leads, the peaks, produced by Kondo effect at equilibrium, split and give rise to
suppressed peaks around the bounds of energy gap of each lead. We also have evaluated the
differential conductance. The Kondo effect in the voltage dependence of differential conductance
shows up as two anomalous peaks around biases $\pm 2\Delta$ and a small peak near zero bias. This
result accords with the experimental observation. With descending the temperature, peaks at biases
$\pm 2\Delta$ ascend and the zero-bias peak falls.

\end{document}